# Entangled-photon decision maker

Nicolas Chauvet[1,2*], David Jegouso[1], Benoît Boulanger[1],

Hayato Saigo[3], Kazuya Okamura[4], Hirokazu Hori[5], Aurélien Drezet[1],

Serge Huant[1], Guillaume Bachelier[1], Makoto Naruse[1,2]

[1] Univ. Grenoble Alpes, CNRS, Institut Néel, 38000 Grenoble, France

[2] Department of Information Physics and Computing, Graduate School of Information Science and Technology, The University of Tokyo, 7-3-1 Hongo, Bunkyo-ku, Tokyo 113-8656, Japan

[3] Nagahama Institute of Bio-Science and Technology, 1266 Tamura, Nagahama, Shiga 526-0829, Japan

[4] Toyohashi University of Technology, 1-1 Hibarigaoka, Tempaku, Toyohashi, Aichi, 441-8580, Japan

[5] Interdisciplinary Graduate School, University of Yamanashi, Takeda, Kofu, Yamanashi 400-8510, Japan

* nicolas_chauvet@ipc.i.u-tokyo.ac.jp

**Abstract**

The competitive multi-armed bandit (CMAB) problem is related to social issues such as maximizing total social benefits while preserving equality among individuals by overcoming conflicts between individual decisions, which could seriously decrease social benefits. The study described herein provides experimental evidence that entangled photons physically resolve the CMAB in the 2-arms 2-players case, maximizing the social rewards while ensuring equality. Moreover, we demonstrated that deception, or outperforming the other player by receiving a greater reward, cannot be accomplished in a polarization-entangled-photon-based system, while deception is achievable in systems based on classical polarization-correlated photons with fixed polarizations. Besides, random polarization-correlated photons have been studied numerically and shown to ensure equality between players and deception prevention as well, although the CMAB maximum performance is reduced as compared with entangled photon experiments. Autonomous alignment schemes for polarization bases were also experimentally demonstrated based only on decision conflict information observed by an individual without communications between players. This study paves a way for collective decision making in uncertain dynamically changing environments based on entangled quantum states, a crucial step toward utilizing quantum systems for intelligent functionalities.

**Introduction**

Unique physical attributes of photons have been intensively studied for information processing to solve computationally demanding problems such as time-series prediction using photonic reservoir computing[1], combinatorial optimization based on coherent Ising machines[2], and deep learning employing nanophotonic circuits for cognition[3]. Decision making is another important branch of research where the objective is to identify decisions that will maximize benefits in dynamically changing uncertain environments[4,5], with direct applications for reinforcement learning. In this context, the multi-armed bandit (MAB) problem is one of the important fundamental problems in decision making, where the objective is to maximize the rewards obtained from multiple slot machines, whose reward probabilities are unknown[4], in contrast with the prisoner problem[6]. To solve the MAB problem, it is necessary to explore better slot machines. However, too much exploration may result in excessive loss, whereas too quick of a decision or insufficient exploration may lead to missing the best machine. We previously successfully solved the MAB problem, by employing excitation transfer via near-field coupling[7], single photons[8,9], and chaotic lasers[10,11]. This type of decision-making problem becomes even more difficult when the number of decision makers, i.e. the number of individuals who join the game or simultaneously play the slot machines, is multiple; then the problem is referred to as a *competitive multi-armed bandit* (CMAB) problem[12,13], which is the focus of the study described herein. In collective decision making, social values are highlighted, such as the maximization of the total social benefits, guarantee of equality among individuals, and so on[12–

[15]. The CMAB problem is important in practical applications ranging from traffic control, where everyone choosing the same road may lead to a traffic jam[16] to resource allocation in infrastructures, such as communications[12,17] where everyone wanting to communicate at the same time leads to congestion for example. A fundamental question asked in the study described herein was whether quantum entanglement[18,19] could bring improvement for reinforcement learning applications[20,21] or resolve the difficulties of the CMAB problem. The usefulness of entangled photons is addressed in the quantum game literature[22–25] regarding resolving Nash equilibrium in non-cooperative games formulated by payoff matrices in game theory[26]. The study described herein was focused on the CMAB problem, which differs from the non-zero-sum game[24] in the sense that the reward in the CMAB problem is not given deterministically, unlike in conventional game theory, but rather probabilistically; thus, one can lose even when the choice is correct, and vice versa. Hence, it is not possible to address the CMAB problem using the payoff matrix formulation alone.

This paper theoretically and experimentally demonstrates the usefulness and superiority of quantum-entangled photons for collective decision making and physically solving the MAB problem on the *social level*, for example, maximizing the total benefits while preserving equality among individuals by overcoming conflicts between individual decisions. Moreover, we demonstrate that deception, or greedily outperforming the other player by trying to receive a greater reward than him, is impossible in a polarization-entangled-photon-based system, while such greedy action is achievable in systems based on classical polarization-correlated photons with fixed polarizations.

Autonomous and dynamic alignment schemes for polarization bases, which are necessary for CMAB applications, are also experimentally demonstrated based only on decision conflict information observed by an individual without communications between players. To the best of our knowledge, such a detailed analysis of the polarization basis, which is also necessary for quantum games, has not been provided elsewhere. In the Discussion, we consider the case of randomly cross-polarized photon pairs and compare its performance with that of entangled photon pairs. The physical limit of photon sources for collective decision making is also discussed. Finally, we study the influence of using a statistical mixture of entanglement-degree-modulated photons and non-entangled photons on both social welfare and individual freedom for players. Although the following discussion is restricted to photonic entangled states, the transposition to any other type of entanglement system is straightforward, giving a broad generality to the present study.

## Decision Making

### System architecture

For the simplest case that preserves the essence of the CMAB problem, we consider two players (called Players 1 and 2 hereafter), each of whom selected one of two slot machines (Machines A and B hereafter), with the goal of maximizing the total social reward. The reward probabilities of Machines A and B are denoted as $P_A$ and $P_B$, respectively. The amount of reward that could be dispensed by each slot machine per play is assumed to be unity even when multiple players choose that same machine. Possibly, the two players make the same decision at the same time, causing conflict between their decisions. In that case, the reward is divided into two halves, which are allocated to the two players. In terms of reward, this penalty is intended to favour collective, i.e. not conflictual, choices with respect to individual interests. From the viewpoint that the two individuals playing the casino act as a *team*, when a player chooses the best slot machine, the other one should select the *other* machine to maximize the sum of their rewards. This example manifests itself as players easily becoming locked in a local minimum due to conflict between their decisions, since everyone wants more rewards and tries to select the higher-reward-probability slot machine, whereas the total team rewards could be increased if they cooperated[12].

As the decision is based on simultaneous photon detection, the "two-player" configuration requires photon pairs to be generated. An overview of the dedicated experimental setup is shown in Fig. 1. It is based on a standard Sagnac loop architecture[27] used to generate the photon states by spontaneous parametric down conversion (SPDC), and analyse them for selecting Machine A or Machine B. The details are described in the Methods section. In the branch corresponding to Player 1, each signal photon goes through a half-wave plate ($HW_1$) and is subjected to a PBS ($PBS_1$). If the photon is detected by the avalanche photodiode corresponding to the horizontally polarized light (APD1), the decision of Player 1 is to choose Machine A, whereas if the photon is detected by the avalanche photodiode corresponding to the vertically polarized light (APD2), then the decision of Player 1 is to choose Machine B. The same hold for player B by exchanging 1 by 3 and 2 by 4. Note that the two slot machines are externally arranged: we emulate the slot machines in a computer using pseudorandom sequences (see Methods for details).

**Decision making by a *single* player**

We start with the *single*-player situation, in which either Player 1 or Player 2 attacks the casino. This is essentially equivalent to the experimental demonstration described in Ref. 8. In this case, it is desirable for the player to choose the higher-reward-probability slot machine, since a larger reward is desired. In order to specify the player action, we now introduce several notations to describe the system. The input photon state for the decision making of Player $i$ ($i=1,2$) is denoted as $|\theta_i\rangle$, $\theta_i$ being the polarization angle; it is delivered by the photon source and is therefore not controllable by the players. The action of Player $i$ is to rotate HW$_i$ by an angle $\theta_{\text{HW}_i}$ which modifies the photon state polarization. The roles of HW$_i$ and PBS$_i$ are given respectively by

$$HW_i|\theta_i\rangle = |2\theta_{\text{HW}_i} - \theta_i\rangle \tag{1}$$

and

$$PBS_i|2\theta_{\text{HW}_i} - \theta_i\rangle = \cos(2\theta_{\text{HW}_i} - \theta_i)|H_i\rangle + \sin(2\theta_{\text{HW}_i} - \theta_i)|V_i\rangle, \tag{2}$$

where $|H_i\rangle$ and $|V_i\rangle$ indicate photon states with horizontal and vertical polarization propagating in orthogonal directions beyond PBS$_i$[28]. Therefore, the probabilities of photon measurement by APD1 and APD2, for example, which determine whether Player 1 decides to select Machine A or Machine B, are given by $\cos^2(2\theta_{\text{HW}_1} - \theta_1)$ and $\sin^2(2\theta_{\text{HW}_1} - \theta_1)$, respectively. Using the tug-of-war principle described in Ref. 8, which is also summarized in the Sec. 1 of Supplementary information, the wave plate angle is controlled toward the higher-reward-probability slot machine. The reward probabilities of Machines A and B are chosen as $P_A = 0.2$ and $P_B = 0.8$, respectively, for the first 50 plays. In the next 50 plays, the reward probabilities are swapped, i.e. $P_A = 0.8$ and $P_B = 0.2$, to emulate a variable environment (Fig. 2a). Therefore, from the standpoint of *individual* players, selecting Machine B is the *correct* decision in the first 50 plays since it is highly likely to provide a greater reward. Likewise, choosing Machine A is correct for the next 50 plays. The adaptive decision making is implemented by updating the wave plate orientation toward the higher-reward-probability slot machine by revising the polarization adjuster (PA) values[8]. First, only Player 1 plays the casino. Specifically, Player 1 conducts 100 consecutive slot plays, and this set of plays is repeated 10 times. The red curve shown in Fig. 2b,i represents the *correct decision ratio* (*CDR*) defined as the ratio of the number of selections of the machine yielding a higher reward probability over the number of trials at cycle *t*. This ratio quickly approaches unity, meaning that Player 1 effectively chooses the higher-reward-probability machine, i.e. Machine B. At cycle 51, the CDR drops due to the flip of the reward probabilities. However, the CDR gradually returns to unity as time elapses, which clearly indicates that Player 1 detects the change in the environment and revises the decision to the higher-reward-probability machine, i.e. Machine A. The red curve in Fig. 2b,ii shows the evolution of the *accumulated reward* averaged over 10 repetitions, which almost linearly increases with time. Its growth is attenuated after cycle 50 due to the reward probability change. Note that the accumulated reward of Player 1 at cycle 100 is about 66.

The blue curves in Figs. 2c,i and 2c,ii show the CDR and accumulated reward when only Player 2 played the slot machines. The behaviour is similar to the case of Player 1. The accumulated reward at cycle 100 is 67, which is almost equivalent to that in the case of Player 1. This finding demonstrates the successful decision making by single players as well as the validity of the strategy adopted for solving the asymmetry between APD collection efficiencies (see Methods).

**Decision making by two *non-cooperative* players: evidence of interest conflict**

Let us now consider the case of 2 Players simultaneously using the casino, using the configuration described in Sec. 2 of the Supplementary Information. Suppose now that both Players 1 and 2 independently play the slot machines; in other words, in a non-cooperative manner. The red and blue curves in Fig. 2d,i are the CDRs of Players 1 and 2, respectively, both of which exhibiting traces similar to those in the single-player cases (Figs. 2b and 2c). Actually, both Players 1 and 2 succeed in finding the higher-reward-probability slot machine over time. However, this result points toward conflicts between their decisions; hence, the accumulated rewards of Players 1 and 2 shown by the red and blue curves, respectively, in Fig. 2d,ii are seriously decreased, i.e. nearly half of those in the single-player cases. The summation of the accumulated rewards of Players 1 and 2, referred to as the *team* reward, is depicted by the green curve in Fig. 2d,ii and is 70.9 at cycle 100, which is only slightly larger than in the single-player cases. Indeed, the *conflict ratio*, which is defined as the number of times that the decisions of Players 1 and 2 are identical over the 10 repetitions, exhibits high values close to unity, as shown by the red curve in Fig. 2d,iii. This result indicates that conflicts between decisions occur very frequently during the slot plays.

**Decision making by cooperative players: polarization-correlated *versus* polarization-entangled photons**

Toward realizing collective decision making, two conditions must be fulfilled. The first method of avoiding conflicts between decisions is to introduce correlations between the two photons, thereby statistically linking the decisions of Players 1 and 2. To this end, we deliver polarization-orthogonal photon pairs denoted by $|\theta_1, \theta_2\rangle$, where

$$\theta_2 = \theta_1 + \pi/2, \tag{3}$$

to the two players as input photon states. In practice, $\theta_1 = 0$ and $\theta_2 = \pi/2$, corresponding to a horizontal polarization for Player 1 and a vertical polarization for Player 2 in the PBS polarization basis. The underlying idea of using orthogonal polarizations is to promote the players to select distinct machines. The actions of the two players are again to rotate the waveplates and to analyse the photon states through the polarization beam splitter, represented by Eqs. (1) and (2). The probability amplitudes of observing photons at one of APD1 and APD2, *and* at one of APD3 and APD4 are as follows:

[APD1 and APD3] $\langle H_1, H_2 | M | \theta_1, \theta_2 \rangle = \cos(2\theta_{HW_1} - \theta_1)\cos(2\theta_{HW_2} - \theta_2)$, (4)

[APD1 and APD4] $\langle H_1, V_2 | M | \theta_1, \theta_2 \rangle = \cos(2\theta_{HW_1} - \theta_1)\sin(2\theta_{HW_2} - \theta_2)$, (5)

[APD2 and APD3] $\langle V_1, H_2 | M | \theta_1, \theta_2 \rangle = \sin(2\theta_{HW_1} - \theta_1)\cos(2\theta_{HW_2} - \theta_2)$, (6)

[APD2 and APD4] $\langle V_1, V_2 | M | \theta_1, \theta_2 \rangle = \sin(2\theta_{HW_1} - \theta_1)\sin(2\theta_{HW_2} - \theta_2)$, (7)

where $M$ denotes the operator describing the action of wave plates and polarization beam splitters. The coincidence of observing photons at APD1 and APD3, and at APD2 and APD4 according to Eqs. (4) and (7), respectively, indicates conflict between the decisions made by Players 1 and 2.

Correlated photon polarization is not a sufficient condition to prevent conflict: in view of both players acting as a team, they must also perform coherent choices. This insufficiency led to the second requirement, namely, the use of correlated wave plate angles. Here, we represent this condition by the rotation of both wave plates by the same amount; that is:

$$\theta_{HW_2} = \theta_{HW_1}.$$ (8)

By subjecting Eqs. (3) and (8) to Eqs. (4) to (7), the probability of conflict between decisions is

$$P_C = \frac{1}{4}\left[1 - \cos\left(8\theta_{HW_1} - 4\theta_1\right)\right],$$ (9)

which can be obtained by summing the squared moduli of Eqs. (4) and (7), while the probability of *no* conflict between decisions is:

$$P_{NC} = 1 - P_C = \frac{1}{4}\left[3 + \cos\left(8\theta_{HW_1} - 4\theta_1\right)\right].$$ (10)

From Eq. (9), $\theta_{HW_1}$ should be configured as $\frac{\theta_1}{2} + \frac{\pi}{4} \times N$, with $N$ being a natural integer to avoid conflict between decisions. Note that this is technically impossible if the photon state $\theta_i$ randomly varies from photon to photon, as it would be the case with a non-polarized classical source of light. Furthermore, even when conflict between decisions is successfully avoided, i.e. by using fixed $\theta_i$ as is the case for signal and idler photons produced by SPDC, the resulting decision is *biased* toward a specific machine, leading to a reward distribution that favours a specific player. It means that the *equality* between the players decreases.

To avoid conflict between decisions, the probability amplitudes in Eqs. (4) and (7) must always vanish, requiring a second contribution exactly cancelling the oscillating sine and cosine terms. To this end, we utilize a coherent superposition of states corresponding to entangled states. Due to the symmetry of the equations, a natural choice is to exchange the roles of $\theta_1$ and $\theta_2$ and to introduce a $\pi$ phase shift, or, in other words, to use a specific entangled state known as the maximally entangled singlet photon state and given by

$$\frac{1}{\sqrt{2}}(|\theta_1,\theta_2\rangle - |\theta_2,\theta_1\rangle), \tag{11}$$

where $\theta_1$ and $\theta_2$ are orthogonal to each other as specified in Eq. (3). Usually, maximally entangled photons are represented by the forms such as $\frac{1}{\sqrt{2}}(|HV\rangle - |VH\rangle)$ or $\frac{1}{\sqrt{2}}(|0,1\rangle - |1,0\rangle)$ unlike Eq. (11). The reason behind the utilization of Eq. (11) is that we keep coherent notations among single photons ($|\theta_1\rangle, |\theta_2\rangle$) and polarizatioin-correlated photons ($|\theta_1,\theta_2\rangle$) introduced above based on the polarization angles of $\theta_1$ and $\theta_2$. Also, by representing polarization angles of input photons, the role of half-wave plate are clearly grasped as discussed below. For such reasons, we introduce the form of entangled photons by Eq. (11).

The probability amplitude originating from the *second* term of Eq. (11) can be derived according to the following equations:

[APD1 and APD3] $-\langle H_1, H_2 | M | \theta_2, \theta_1 \rangle = -\cos(2\theta_{HW_1} - \theta_2)\cos(2\theta_{HW_2} - \theta_1)$, (12)

[APD1 and APD4] $-\langle H_1, V_2 | M | \theta_2, \theta_1 \rangle = -\cos(2\theta_{HW_1} - \theta_2)\sin(2\theta_{HW_2} - \theta_1)$, (13)

[APD2 and APD3] $-\langle V_1, H_2 | M | \theta_2, \theta_1 \rangle = -\sin(2\theta_{HW_1} - \theta_2)\cos(2\theta_{HW_2} - \theta_1)$, (14)

[APD2 and APD4] $-\langle V_1, V_2 | M | \theta_2, \theta_1 \rangle = -\sin(2\theta_{HW_1} - \theta_2)\sin(2\theta_{HW_2} - \theta_1)$. (15)

The probability of photodetection at both APD1 and APD3, meaning that both Players 1 and 2 select Machine A, is then given by the squared modulus of the *coherent* sum of Eqs. (4) and (11), both multiplied by $1/\sqrt{2}$, which leads to:

$$P_{1\&3} = \frac{1}{2}\sin^2\left[2\left(\theta_{HW_1} - \theta_{HW_2}\right)\right] = P_{2\&4}. \quad (16)$$

This probability *always* yields zero *regardless* of the values of $\theta_i$ and $\theta_{HW_i}$ as long as the conditions of Eqs. (3) and (8) apply. Likewise, the probability that both Players 1 and 2 select Machine B is given by Eq. (16) and therefore is also *always zero*. Thus, conflicts between decisions *never* occur, leading to the maximum overall social reward. Conversely, the probability of observing photons at APD1 and APD4 or at APD2 and APD3 can be expressed as

$$P_{1\&4} = P_{2\&3} = \frac{1}{2}\cos^2\left[2\left(\theta_{HW_1} - \theta_{HW_2}\right)\right], \quad (17)$$

which is *always* 0.5 when Eqs. (3) and (8) are satisfied. Thus, both players have equal opportunities to select each slot machine, which is the foundation of the equality provided by polarization-entangled photons.

Figure 3 characterizes the details of the collective decision-making performance with respect to the polarization basis. Figures 3a,i and 3b,i depict the accumulated reward at cycle 100 as a function of the common orientation of the half-wave plates, which corresponds to the common polarization basis, regarding the decision making based on polarization-correlated and polarization-entangled photon pairs, respectively. The red squares, blue diamonds, and green circles correspond to the rewards received by Player 1, Player 2, and the team, respectively.

The team reward is very high, about 100, when the polarization is 0° and 90°, even in the case of correlated photon pairs. For example, a correlated photon pair given by $|H_1,V_2\rangle$ is a photon pair corresponding to the polarization angle $\theta_1 = 0°$. In this case, Player 1 always detects a $|H_1\rangle$ photon, leading to the decision to choose Machine A, whereas Player 2 always detects a $|V_2\rangle$ photon. It indicates that the decision is to select Machine B. Therefore, from the viewpoint of *correct* decision making, Player 2 achieves a higher CDR in the first 50 cycles and a lower CDR in the second 50 cycles due to the reward probability flipping, as clearly demonstrated in Fig. 3a,ii[0]. A similar tendency is observable with the input photon pair described by $|V_1,H_2\rangle$ that corresponds to the polarization angle $\theta_1 = 90°$, as shown in Fig. 3a,ii[90]. The conflict ratio remains lower in these specific polarization cases than others, as shown by the red squares in Fig. 3a,iii.

However, in terms of *equality* (or *fairness*) this scenario is highly inefficient since a player can select either machine. Indeed, with 0° polarization, only Player 2 earns a greater reward in the first 50 plays, and the imbalance between Players 1 and 2 is significant. More specifically, the equality depicted by the diamonds in Fig. 3a,iii is significantly decreased; it is defined as the average ratio between the numbers of times that the higher-reward-probability machine was selected by Players 1 and 2. The exact definition of equality is provided in Sec. 3 of Supplementary information.

On the contrary, with the use of entangled photons, the team reward always reaches the theoretical maximum (100) regardless of the common polarization basis, as shown in Fig. 3b,i. It is due to the maximally entangled state that is invariant upon rotation of the basis, provided that the bases are the same for both players. The CDRs of Players 1 and 2 always randomly fluctuate around 0.5, as shown in Fig. 3b,ii. This fluctuation agrees with the fact that nearly identical rewards were received by Players 1 and 2, as can be seen in Fig. 3b,i. The conflict rate, shown by the red squares in Fig. 3b,iii, is always small regardless of the polarization basis, though non-zero due to experimental imperfections. Finally, the equality remains always high for all of the common polarization bases, clearly showing that the entangled states yielded results superior to those achievable using the correlated states in terms of all of the investigated social properties.

To summarize the figures of merit of all of the decision-making strategies, the total rewards resulting from using single-photon decision making for single players and non-cooperative and collective decision making are compared in Fig. 4. The orange and green bars depict the experimental and simulation results, respectively, which agree well throughout the experiments. For the two-player games, the experimentally obtained rewards of the individual players are also shown: red for Player 1 and purple for Player 2. The diagonal and vertical stripes areas indicate the rewards accumulated during the first 50 and second 50 plays, respectively, emphasizing the effective equality or inequality between the two players. The error bars show the maximum and minimum observed values. Clearly, the maximum team reward is achieved by using entangled photons. Furthermore, the individual rewards in the entangled photon case are higher than those in the case of two non-cooperative players, indicating that non-conflict and equal opportunities not only lead to the social maximum, but also benefit the individual players.

In order to check the sensitivity of the previous results on the reward probabilities $P_A$ and $P_B$, the latter are changed to $P_A = 0.4$ and $P_B = 0.6$. The data is shown in Sec. 4 and Fig. S2 in the Supplementary information. It appears that finding the higher-reward-probability machine is more difficult in that case due to the smaller difference between the reward probabilities than in the former cases. As a consequence, the total reward is substantially lower. These differences are due to the longer time needed to reach stable selection of the higher-reward-probability machine. On the contrary, with correlated and entangled photons, the team reward does not change, and the entangled photons again provide the maximum total reward. This finding clearly demonstrates that collective decision making based on entangled photons ensures that the social maximum reward will be achieved regardless of the difficulty of the given problem. This has strong implications in terms of allocation resources as for example in network communications as the maximized efficiency is ensured whatever the actual qualities of the two channels which may fluctuate in time.

## Deception or greedy action

An important condition for establishing the social maximum in the CMAB solution by using polarization-entangled photons is sharing of the polarization basis among the players. However, one of the players could have his/her basis misaligned, for instance upon trying to increase his/her own reward on the detriment of the other player's – an action called deception, or greedy action. In the following, we theoretically investigate how polarization-correlated and polarization-entangled photons allow or inhibit "deception", characterized by $\theta_{HW2} \neq \theta_{HW1}$ (i.e. Eq. (8) no longer holds).

### Greedy action in *correlated-photon*-based systems

Using polarization-correlated photon pairs characterized by Eqs. (4)–(7), the expected reward received by Player 1 in a single play is

$$E[R_1] = P_A \times \cos^2(2\theta_{HW_1} - \theta_1)\sin^2(2\theta_{HW_1} - \theta_1) + P_A \times \cos^2(2\theta_{HW_2} - \theta_1)\cos^2(2\theta_{HW_2} - \theta_1) \times 0.5 \\ + P_B \times \sin^2(2\theta_{HW_1} - \theta_1)\sin^2(2\theta_{HW_2} - \theta_1) \times 0.5 + P_B \times \cos^2(2\theta_{HW_2} - \theta_1)\sin^2(2\theta_{HW_1} - \theta_1) \quad (18)$$

The multiplication factor 0.5 in the second and third terms on the right-hand side indicates that the reward is halved due to the conflict between decisions. Eq. (18) can be reduced to

$$E[R_1] = \frac{3}{8}(P_A + P_B) + \frac{1}{16}(P_A + P_B)\cos[4(\theta_{HW_1} - \theta_{HW_2})] \\ + \frac{1}{16}(P_A + P_B)\cos[4(\theta_{HW_1} - \theta_{HW_2} - \theta_1)] \\ + \frac{3}{8}(P_A - P_B)\cos(4\theta_{HW_1} - 2\theta_1) + \frac{1}{8}(P_A - P_B)\cos(4\theta_{HW_1} - 2\theta_1) \quad (19)$$

Likewise, the expected reward received by Player 2 is

$$E[R_2] = \frac{3}{8}(P_A + P_B) + \frac{1}{16}(P_A + P_B)\cos\left[4(\theta_{HW_1} - \theta_{HW_2})\right]$$
$$+ \frac{1}{16}(P_A + P_B)\cos\left[4(\theta_{HW_1} - \theta_{HW_2} - \theta_1)\right] \qquad (20)$$
$$+ \frac{3}{8}(P_B - P_A)\cos(4\theta_{HW_1} - 2\theta_1) + \frac{1}{8}(P_B - P_A)\cos(4\theta_{HW_1} - 2\theta_1)$$

From Eqs. (19) and (20), the expected amount by which the reward of Player 2 exceeds that of Player 1 is given by:

$$E[R_2] - E[R_1] = \frac{1}{2}(P_B - P_A)\left[\cos(4\theta_{HW_1} - 2\theta_1) + \cos(4\theta_{HW_2} - 2\theta_1)\right]. \qquad (21)$$

It means that the expected reward can be biased toward a particular player depending on the difference between the polarization bases and the incoming photon polarization. In addition, this characteristic implies that no matter what $\theta_1$ and $\theta_{HW_1}$ are, it is *possible* for Player 2 to receive a reward greater than (or at least equal to) that received by Player 1 by configuring

$$\theta_{HW_2} = \frac{1}{2}\theta_1 + \frac{N\pi}{2}, \qquad (22)$$

where $N$ is a natural integer. Thus, *deception*, or greedy action by a player in the system to gain a greater reward than the other, is generally *achievable* when the system is governed by correlated photons.

**Greedy action in *polarization-entangled-photon*-based systems**

Following the same procedure for polarization entangled photons, the expected reward received by Player 1 for a single play is:

$$E[\mathrm{R}_1] = P_A \times \frac{1}{2}\sin^2\left[2\left(\theta_{\mathrm{HW}_1} - \theta_{\mathrm{HW}_2}\right)\right] + P_A \times \frac{1}{2}\cos^2\left[2\left(\theta_{\mathrm{HW}_1} - \theta_{\mathrm{HW}_2}\right)\right] \times 0.5$$
$$+ P_B \times \frac{1}{2}\cos^2\left[2\left(\theta_{\mathrm{HW}_1} - \theta_{\mathrm{HW}_2}\right)\right] \times 0.5 + P_B \times \frac{1}{2}\sin^2\left[2\left(\theta_{\mathrm{HW}_1} - \theta_{\mathrm{HW}_2}\right)\right] \quad (23)$$

or equivalently:

$$E[\mathrm{R}_1] = \frac{3}{8}(P_A + P_B) + \frac{1}{8}(P_A + P_B)\cos\left[4(\theta_{\mathrm{HW}_1} - \theta_{\mathrm{HW}_2})\right]. \quad (24)$$

The expected amount of reward received by Player 2 also results in Eq. (24). That is, no matter how Player 2 configures $\theta_{\mathrm{HW}_2}$, the rewards allocated to Players 1 and 2 are the same. Thus, even if Player 2 knows the higher-reward-probability machine and can rotate the wave plate with the intention of receiving a greater reward, such deception is *impossible* if the system is governed by polarization-entangled photons. Moreover, the expected total reward received by Players 1 and 2, given by two times Eq. (24), i.e.:

$$E[\mathrm{R}_{\mathrm{TEAM}}] = \frac{3}{4}(P_A + P_B) + \frac{1}{4}(P_A + P_B)\cos\left[4(\theta_{\mathrm{HW}_1} - \theta_{\mathrm{HW}_2})\right] \quad (25)$$

is *less than* its maximum value, given by $P_A + P_B$, if the half-wave plate alignment is disrupted ($\theta_{\mathrm{HW}_1} \neq \theta_{\mathrm{HW}_2}$) unless $\theta_{\mathrm{HW}_2} = \theta_{\mathrm{HW}_1} + N \times \pi/2$ with $N$ being a natural integer. That is, in addition to the inhibition of deception, the total social benefits are decreased if a selfish action is performed by one of the players.

**Experimental investigation of Greedy action**

We now investigate a scenario in which one of the players (here Player 2) is *greedy*, and tries to *deceive* the other player to obtain a greater reward by rotating his half-wave plate in the direction of the higher-reward-probability machine. The orientation of the half-wave plate is controlled toward the higher-reward-probability slot machine by revising the polarization control (PC) value[8]. The PC value is limited to a maximum and minimum of 10 and −10. Essentially, larger (positively large) and smaller (negatively large) PC values indicate that the half-wave plate is rotated so that the polarization of the photon is toward the horizontal and vertical directions, respectively. (The details of the PC values are described in Sec. 1 of Supplementary information.)

Figure 5 summarizes the total reward obtained with polarization-entangled photons at cycle 100 for Player 1, Player 2, and the team. The error bars indicate the maximum and minimum rewards. Clearly, the preservation of equality between players and decrease of the team reward obtained for polarization-entangled photons (Fig. 5a) agree with the theoretical analysis. In contrast, for polarization-correlated photons (Fig. 5b), Player 2 achieves deception through this greedy action, thereby destroying the equality, with almost no effect on the total team reward. This is the worst configuration, as selfish action only benefits to its author and not at least indirectly to the team.

**Autonomous polarization-basis alignment**

As discussed above, polarization-basis alignment between the players is crucial to realize the maximal social benefits. However, the optical system may suffer from certain environmental disturbances during the decision-making operations that degrade its performance. Therefore, on-line calibration that does not interrupt the decision-making operation is important and should be performed by the players and not by the photon provider. Here we discuss the resolution of these issues by two different methods considering that the goal is to configure $\theta_{HW_2}$ with respect to the unknown $\theta_{HW_1}$ using an adaptation algorithm.

**Assumption I: no prior information about the polarization basis**

We first investigate the possibility of aligning the polarization bases without any prior information, exploiting the fact that a half-reward event indicates conflict between decisions. Simultaneously, when the polarization bases are aligned, the probability of conflict between decisions, i.e. $\sin^2\left[2\left(\theta_{HW_1} - \theta_{HW_2}\right)\right]$, is zero. Therefore, an alignment strategy is as follows.

> **[K-0]** If the receipt of a half-reward is observed, update $\theta_{HW_2}$ by $\theta_{HW_2} + \Delta_a$.

Here, $\Delta_a$ is a constant employed to change $\theta_{HW_2}$ gradually. If $\Delta_a$ is sufficiently small, by repeating **[K-0]**, the difference between the half-wave plate angles $\theta_{HW_1} - \theta_{HW_2}$ should eventually become small; hence, the probability of conflict between decisions should decrease.

In the experimental demonstration, $\theta_{HW_2}$ is initially –22.5° and $\Delta_a$ is 12.5°. $\theta_{HW_2}$ should be made equal to $\theta_{HW_1}$, which is 0°. The evolution of $\theta_{HW_2}$ in each sequence is shown by the blue curves in Fig. 6a, while the target angle $\theta_{HW_1} = 0°$ is depicted by the red lines. After applying **[K-0]** twice, $\theta_{HW_2}$ increases by +25°; hence, $\theta_{HW_2}$ becomes 2.5°, which is sufficiently close to 0°. Even though $\theta_{HW_2}$ cannot be exactly zero, it is evident from the evolution of $\theta_{HW_2}$ shown in Fig. 6a that $\theta_{HW_2}$ *passes though* the target angle and continues increasing. That is, even when $\theta_{HW_2} - \theta_{HW_1}$ is very small, conflict between decisions cannot be perfectly avoided due to the imperfections of the experimental system (see Fig. S1 in Supplementary information).

**Assumption II: no prior information about the polarization basis but memorization of conflict allowed**

To prevent such escape from the recalibrated angle due to error signals, one idea is to take the history into account. The revised calibration strategy is as follows.

> **[K-M]** If conflicts between decisions have *not* been detected in the past $M$ plays, the detection of conflict between decisions in the current play is *discarded*. If there were $K$ occurrences of conflicts between decisions in the past $M$ plays, *and* the current play yields a conflict between decisions, then update $\theta_{HW_2}$ by $\theta_{HW_2} + \Delta_a$ and register the occurrence of a conflict between decisions in the memory of the $M$ most recent plays.

In the experimental implementation, $M$ and $K$ are 5 and 1, respectively, which we call **[1-5]**, while $\Delta_a = 11.25°$. As shown by the blue curves in Fig. 6b, the half-wave plate angle successfully approaches the target angle. With $\Delta_a = 11.25°$, two position updates via **[K-M]** perfectly resolve the initially imposed misalignment ($-22.5°$), so that the effect of memorizing past events is clear.

Indeed, although it is rare, the case of Fig. 6b shows the half-wave plate angle still passing through the target angle. By more severely restricting the condition of rotating $\theta_{HW_2}$ in increasing $M$, the robustness against errors increases. As shown by the blue curves in Fig. 6c where $M$ and $K$ are 10 and 1, respectively, which is referred to as **[1-10]**, the event of passing through the target angle is avoided; however, the adaptation is very slow. Specifically, too large of a memory ($M = 10$, **[1-10]**) provides robustness against errors but results in very slow responses, whereas no memory ($M = 0$, **[1-0]**)) yields a fast response but reactions that are too sensitive to error signals. A moderate parameter choice (**[1-5]**) resolves both the error tolerance and alignment speed issues. The green, red, and blue curves in Fig. 6d summarize the evolution of the accumulated team rewards based on the **[1-0]**, **[1-5]**, and **[1-10]** calibration rules, respectively, where **[1-5]** is optimal for maximizing the total team rewards.

## Discussion

As demonstrated herein, entangled photons enable the achievement of maximum social rewards, equality among individuals, and prevention of selfish actions in communities when solving the CMAB problem. Clear differences between polarization-correlated and polarization-entangled photons are also observed.

The correlated photon pairs mentioned so far do not share the same behaviour as the polarization-entangled photon pairs. In particular, photons from polarization-correlated pairs always have the same input polarization with this description, while entangled photons with state defined as $\frac{1}{\sqrt{2}}(|HV\rangle - |VH\rangle)$ do not have a fixed one. In that sense, a closer equivalence between correlated and entangled photon pairs is a series of photon pairs cross-polarized along random direction, each with a state of the form $|\theta, \theta + \pi/2\rangle$, with $\theta \in [0, 2pi[$ taking random value for each pair. In this way, any given player still has equal probability on average to select one of the two machines whatever its waveplate angle is, though the relative angle values between players' polarization bases will tune the conflict rate and thus influence the total reward.

Section 5 of Supplementary Information is dedicated to this case of study, from which three main observations can be made. The first is that individual and total rewards only depend on the relative angle between polarization bases, as is the case with entangled photons. Secondly, maximum total reward obtained with identical measurement bases is 12% lower than for entangled photon pairs. Finally, no deception strategy is able to make a player earn more reward for himself only, as for the entangled photons case.

To summarize, such a system based on randomly cross-polarized photon pairs show the same dynamics at play with entangled photon pairs: individual and total rewards depend only on relative angles between players' polarization bases, such that the action of a single player is sufficient to improve or reduce both players' outcome. However, randomly cross-polarized photon pairs present weaker variations in comparison with entangled photon pairs, including lower maximum reward (88 compared with almost 100) and higher minimum reward when bases are at 90 degrees from each other (62 against 50). This kind of resource may then be of interest for applications where lower sensitivity to perturbations is needed, whereas entangled photon pairs are more interesting for maximum performance in low-perturbation conditions.

Now we consider the physical limit of photon sources in collective decision making. As far as the action of the players corresponds to the rotation of the waveplates, the players are fully independent from the photon source. There is no prior relation between the waveplate angles. However, the final decisions of the players depend on the specific photon states from the photon source, being either fixed polarization-correlated or polarization-entangled states. From this viewpoint, the photon source does influence players' potential decisions. As an example, let us remind the situation where the same waveplate configuration is used for both players (with no rotation for simplicity) in the section of *Decision making* discussed above. For polarization-correlated photons, Player 1 always selects Machine A while Player 2 selects Machine B all the time, introducing a bias if the two machines do not have the same reward probability. For polarization-entangled photons, Player 1 and Player 2 *randomly* select Machine A and Machine B, ensuring equality. Namely, individual decisions cannot be specified by two polarization-entangled photons.

Another interesting feature of quantum entanglement is the robustness against third-party attack or source alteration. Indeed, as has been studied and shown for quantum key distribution with polarization entangled photon pair transmission through optical fibers[29,30], protocols can be elaborated to be able to detect any eavesdropping attack or alteration of the polarization entanglement of the photon pair source. These rely on tests where both players randomly rotate their waveplate at every measurement, then communicate through public channel which rotation angle they used at a given try and verify whether the conflict rate was below a certain limit for a given relative angle between their measurement bases: if the source is indeed sending entangled photon pairs and no one is eavesdropping, conflict rate should fall to almost zero, whereas a compromised source and/or communication channel would necessarily increase the conflict ratio measured by the players. This aspect may be of interest for applications such as sensitive resource allocation.

It is worth stressing that our work exploits quantum entanglement as a fundamental resource for problems involving multi-armed bandits in competitive scenarios. While our proof-of-principle experiments are based on two-photon entanglement, it should be emphasized that this is not a fundamental limitation: regarding photonic states, one could alternatively exploit single-photon entanglement with vacuum in a state like $|0\rangle_A |1\rangle_B - |1\rangle_A |0\rangle_B$ where 0 and 1 denote the number of photons in modes A and B (as used with beam splitters for example), respectively. Therefore, a single photon could in principle do the same job as our two photons. However, working with vacuum states is not that easy: this would ultimately require certain homodyne measurements involving local oscillators interfering with single photons in order to develop unambiguous tests[31] for examples.

If entangled photons turn out to be competitive in terms of social efficiency and equality, the *freedom* of simultaneous decision of Players 1 and 2 is indeed completely ruled out by the strong authority imposed here by the probability properties of entangled states: whatever the angle of a player's waveplate, the selection probability for a given machine remains ½. In contrast, with two non-cooperative players using single photons, although the total team reward is very poor because of the conflicts between decisions, the *freedom* of choosing machines is fully guaranteed, since a given machine selection probability follows a Malus law with respect to the waveplate angle. A *mixture* of (i) social decision making by using an entangled-photon-based decision maker for efficiency and equality within a team, and (ii) individualistic decision making by using a conventional single-photon-based decision maker[8,9] for freedom, is an interesting and important topic for future study, especially in dynamically changing uncertain environments. Simultaneously, the conflict-avoidance nature of entangled photons may accelerate the exploration phase in finding higher-reward-probability selections among many alternatives, which is another topic requiring future research.

Social decision making *and* individualistic decisions may be weighted through the *modulation* of the degree of entanglement[32] in the following form:

$$a|\theta_1, \theta_2\rangle + be^{i\phi}|\theta_2, \theta_1\rangle. \tag{26}$$

The parameters $a$, $b$, and $\phi$ are real numbers, so that an intended social metric is realized, rather than just maximally entangling the photons as done in this study according to Eq. (7). This can be achieved by adjusting $HW_E$ and $QW_E$. A general mathematical formalism, including category theoretic approach[33], would facilitate the understanding of complex interdependencies of the entities.

Finally, the scalability of entangled-photon-based decision makers is another fundamental topic in view of many practical applications. It is indeed technologically challenging to realize entanglement among many photons[34]. The issue of scalability could be addressed by employing for example novel material systems[35–37] or integrated photonic circuits[38]. It could also be addressed by considering entangled photons combined with a certain coding strategy in order to process many bits of information in a time-multiplexed manner[11]. Hence, our pioneering results are anticipated to stimulate concrete implementation of entangled-photon (or more generally entangled-excitation)-based quantum decision makers.

**Conclusion**

We have theoretically and experimentally demonstrated that entangled photons efficiently resolve the CMAB problem so that the total social reward is maximized, and social equality is accomplished, while also preventing deceptive or greedy actions. In solving competitive two-armed bandit problems, two independent players using polarized-single-photon-based decision making find the higher-reward-probability machine, but the total reward is seriously decreased due to the conflicts of interest. Fixed polarization-correlated photon pairs are useful, to some extent, for deriving non-conflicting decisions, providing freedom of choice for players, and obtaining a greater total reward, but they cannot eliminate conflicts between decisions perfectly. Moreover, this method has difficulty to provide equality. In contrast, entangled photons both enable conflicts between decisions to be avoided and the theoretical maximum total reward to be obtained, while guaranteeing equality regardless of the players' polarization bases. By highlighting the polarization-basis requirement for maximum performance with entangled photons, we have investigated the issue of polarization and value alignment in decision making based on polarization-entangled photons. If polarization-entangled-photon-based decision making is employed, we find that deception, or preventing the other player receiving a greater reward by performing greedy actions in the two-armed bandit problem, is impossible thanks to the physical properties of the polarization dependencies derived by quantum superposition of states. In other words, the reward is always equally shared on average among the players. Furthermore, the total common and individual rewards are decreased by greedy action in such a system, such that autonomous alignment schemes based only on interest-conflict information were demonstrated, which can also be used to verify the integrity of the photon pair source and the communication channels. On the contrary, deception is achievable when the decisions are based on fixed polarization-correlated photons. Additionally, we have shown that deception prevention and guaranteeing equality between players is also achievable by using randomly cross-polarized photon pairs, at the cost of a lower maximum achievable reward and a lower sensitivity to misalignment between polarization bases. Entangled-photon-based system are then more interesting for applications where maximum common performance is required or conflicts must be avoided, whereas randomly cross-polarized-photon-based systems can be of interest if stability and lower sensitivity to perturbations are to be privileged.

The present work hence demonstrated that quantum entanglement, as verified with polarization entangled photon pairs, can be a powerful resource for achieving social maximum benefits as well as addressing key features such as preventing greedy actions when solving the CMAB problem. These features are the foundations of important applications, such as secured allocation of precious resources like energy or frequency bands in communication in the age of artificial intelligence.

**METHODS**

We describe here the experimental setup used to generate photon pairs, as well as the emulation system for the slot machines. The output of an excitation laser passed through a polarizer, a half-wave plate (denoted $HW_E$), a quarter-wave plate ($QW_E$), and a dichromic mirror (D), and was incident upon a polarization beam splitter ($PBS_L$) shown in Fig. 1. The horizontally and vertically polarized components of the incoming light travelled clockwise and anti-clockwise, respectively, through a Sagnac loop containing a half-wave plate ($HW_L$) and type II quasi-phase-matched periodically poled KTiOPO$_4$ (PPKTP) nonlinear crystal (Cr), where spontaneous parametric down conversion (SPDC) was induced[39]. The entanglement of orthogonally polarized photons was generated in the $PBS_L$, where the two paths were recombined. The signal and idler photons corresponded to the outgoing components from the $PBS_L$; the signal photons were directed into the branch for the decision making of Player 1, whereas the idler photons travelled to the branch for the decision making of Player 2. Note that the signal and idler photons had distinct wavelengths and were spectrally selected to avoid contamination, which would have affected the final choices of the players.

For entangled photon generation, it was necessary for $HW_E$ and $QW_E$ to be installed properly to satisfy the condition of generating SPDC equally through both optical paths of the Sagnac loop[27]. Thus, *classical*, which means not entangled, *polarization-correlated* photon pairs could also be generated easily by orienting the wave plates so that only the horizontally or vertically polarized component of the excitation laser was incident and travelled through the Sagnac loop either clockwise or anti-clockwise. In addition to the benefits of the superior stability of generating SPDC by the Sagnac loop system[27], the difference between polarization-correlated and polarization-entangled photon pairs could easily be investigated using the same experimental architecture.

A schematic diagram of the experimental setup is shown in Fig. 1. A fibre-pigtailed, diode-pumped, solid-state laser (Obis, 405 FP) operated at a wavelength of 404 nm with an output power of 100 mW supplied excitation light through a quarter-wave plate ($QW_E$) (Thorlabs, WPQ05M-405) and half-wave plate ($HW_E$) (Thorlabs, WPH05M-405) into a PPKTP crystal (Raicol, type-II colinear SPDC cut) in a polarization Sagnac loop built by a polarization beam splitter ($PBS_L$) (OptoSigma, PBSW-12.7-3/7) and half-wave plate ($HW_L$) (Thorlabs, AHWP05M-600)[S1]. The PPKTP crystal was mounted on a Peltier cooler (Raicol, Peltier controller) to hold the temperature at 313 K. The generated signal light was directed into the branch of Player 1 via a dichroic mirror (Thorlabs, BS011), while the idler light was sent to the branch of Player 2. Due to the limitations of the optical bench, 5-m-long optical fibers (Thorlabs, P1-780A-FC-5) were inserted for both branches, followed by half-wave plates ($HW_1$ and $HW_2$) (Thorlabs, WPH05M-808). In the single-player and two-non-cooperative-player cases, polarizers (P*) (Thorlabs, LPNIR050-MP2) were used. The signal and idler light were then separately subjected to a grating installed in a spectrometer (Roper Scientific, SP-2155 Monochromator) to obtain 805 nm and 812 nm light for the signal and idler, respectively. The signal light was incident upon $PBS_1$ (Thorlabs, PBS251) and detected by either APD1 or APD2 (Excelitas, SPCM-AQRH-16). The idler light went to $PBS_2$ (Thorlabs, PBS252) and was detected by either APD3 or APD4. The photon arrival time were evaluated using a 100-ps-bin-size multiple-event time digitizer (time-to-digital converter) (FAST ComTec, MCS6A), which was connected to a host computer (HP, Z400) with an Intel Xeon CPU (2.67 GHz), OS Windows 7 professional 64 bit. Three half-wave plates ($HW_E$, $HW_1$, and $HW_2$) and a quarter-wave plate ($QW_E$) were mounted on motorized rotary positioners (Thorlabs, PRM1Z8) driven via DC servomotors and controlled by the host computer. LabVIEW (version 2012) was used to control the experimental system, including the slot machine emulation.

The slot machines were emulated in the host computer using pseudorandom numbers ranging from 0 to 1. If the random number was smaller than the reward probability of Machine A ($P_A$), a reward was dispensed. The same mechanism applied for Machine B.

The details of the following materials are shown in the Supplementary information.

1. Single-player and two-non-cooperative-player decision-making strategies

2. Implementation of collective decision making

3. Definition of equality

4. Dependence of total rewards on casino setting

5. Randomly cross-polarized photon pairs

## DATA AVAILABILITY

Data used in this study is available upon reasonable request to the corresponding author.

## ACKNOWLEDGEMENTS

This work was supported in part by the CREST project (JPMJCR17N2) funded by the Japan Science and Technology Agency, the Core-to-Core Program A. Advanced Research Networks and Grants-in-Aid for Scientific Research (A) (JP17H01277) funded by the Japan Society for the Promotion of Science and Agence Nationale de la Recherche, France, through the TWIN project (Grant No. ANR-14-CE26-0001-01-TWIN) and Placore project (Grant No. ANR-13-BS10-0007-PlaCoRe), the Université Grenoble Alpes, France, through the Chaire IUA award granted to G.B. and the one-month invited Professorship of M.N. We also acknowledge the Ph.D. grant to N.C. from the Laboratoire d'excellence LANEF in Grenoble (ANR-10-LABX-51-01). The authors thank the valuable contribution of Yannick Sonnefraud who had initiated the entangled photon experiment in the Grenoble group.

**AUTHOR CONTRIBUTIONS**

M.N., H.H., A.D., S.H., and G.B. directed the project. M.N. and G.B. designed the system architecture. N.C., D.J., B.B., and G.B. designed and implemented the entangled photon system. N.C., D.J., M.N., and G.B. conducted the optical experiments. B.B., H.S., K.O., H.H., A.D, S.H., and G.B. investigated theoretical fundamentals. M.N., N.C., and G.B. analysed the data. M.N., N.C., B.B., S.H., and G.B. wrote the paper.

**COMPETING INTERESTS**

The authors declare no financial or non-financial competing interests.

**Correspondence and requests for materials** should be addressed to N.C.

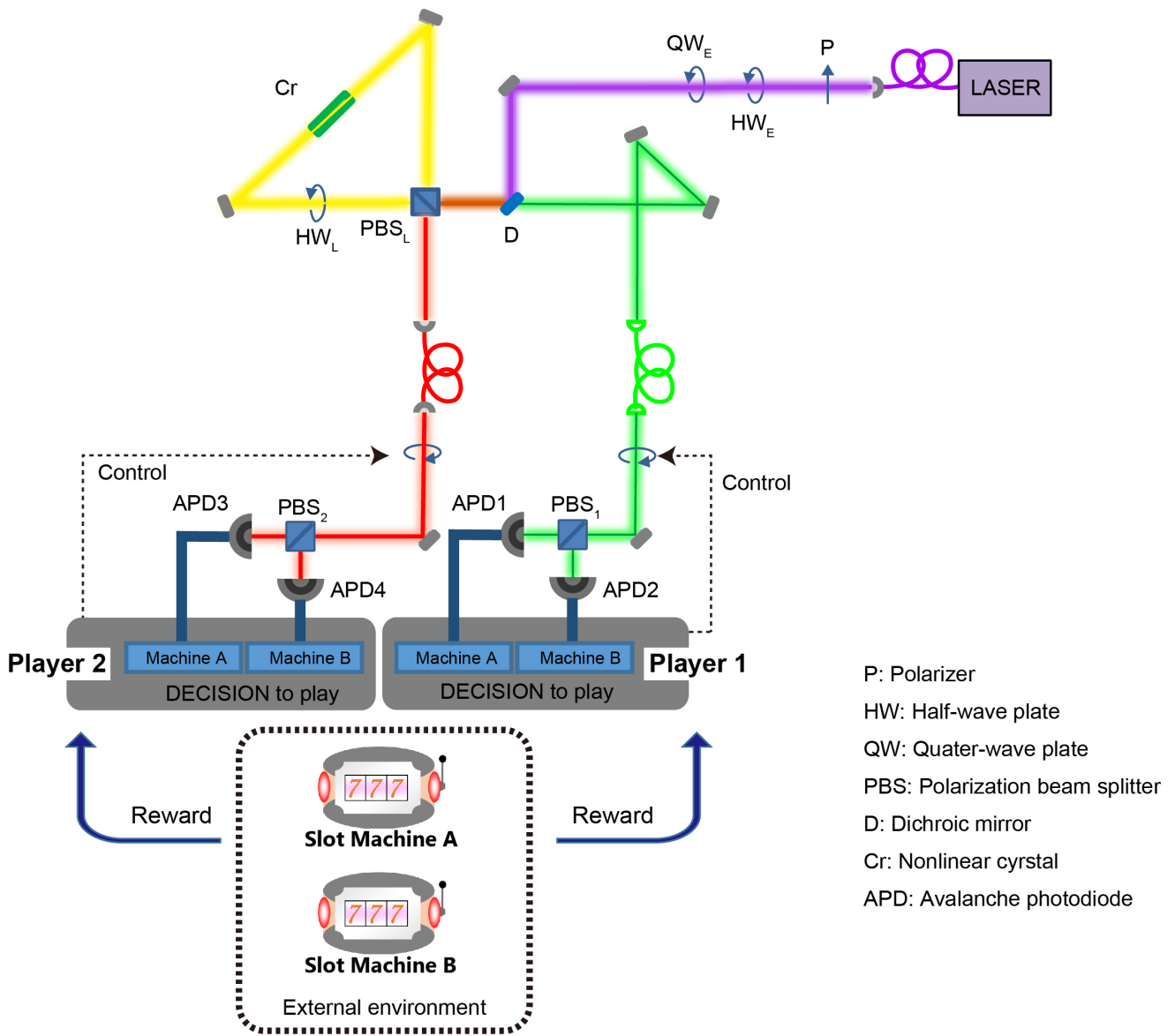

**Fig. 1 | Experimental architecture for solving the CMAB problem using entangled photons.** Spontaneous parametric down-conversion is induced in a nonlinear PPKTP crystal inserted in a Sagnac interferometer architecture. The signal light is used for the decision of Player 1, while the idler light is used for that of Player 2[27]. By configuring the half-wave and quarter-wave plates in front of the excitation laser, polarization-correlated or polarization-entangled photon pairs can be equally generated. The two slot machines (Machines A and B) are external environments, which are emulated in the host computer.

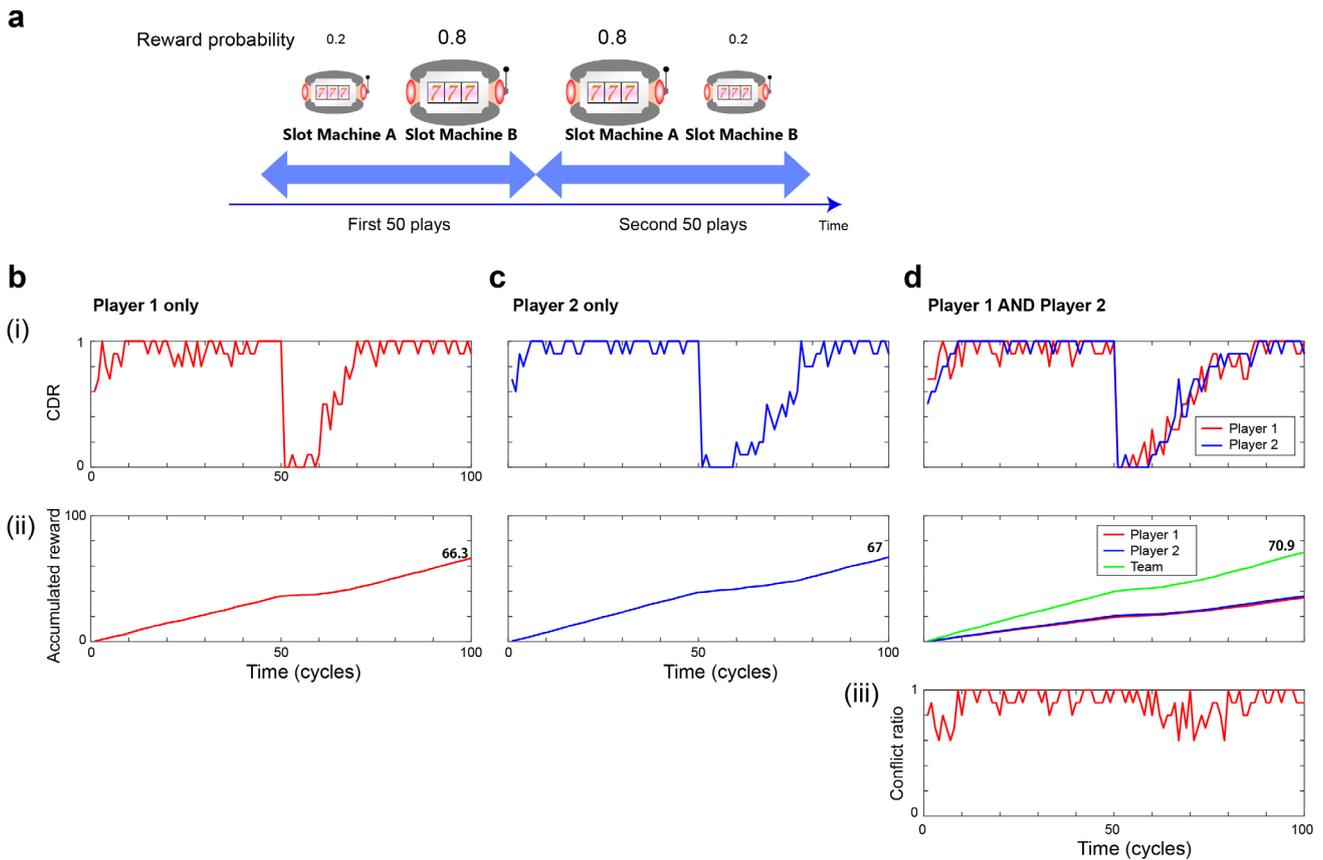

**Fig. 2 | Decision making by a single player and two non-cooperative players. a** Schematic illustration of the casino setting: the reward probability of Machine B is higher ($P_B = 0.8$) than that of Machine A ($P_A = 0.2$) in the first 50 plays, whereas that of Machine A is higher ($P_A = 0.8$) than that of Machine B ($P_B = 0.2$) in the second 50 plays. **b** Decision making when only Player 1 plays the casino. (i) The *CDR*, which is the ratio of choosing the higher-reward-probability slot machine over the number of trials, adaptively approaches unity, meaning that Player 1 is making *good* decisions. (ii) The *accumulated reward* linearly increases over time. **c** Decision making when only Player 2 plays the machines. **d** Decision making when both Players 1 and 2 play the machines. The CDRs of both players adaptively approach unity; that is, both players choose the higher-reward-probability machine. However, making the same decision causes conflict between their decisions, limiting the rewards for each of the players as well as the team rewards (ii). (iii) The *conflict ratio*, which is the ratio of the occurrence of identical decisions by the two players over the number of trials.

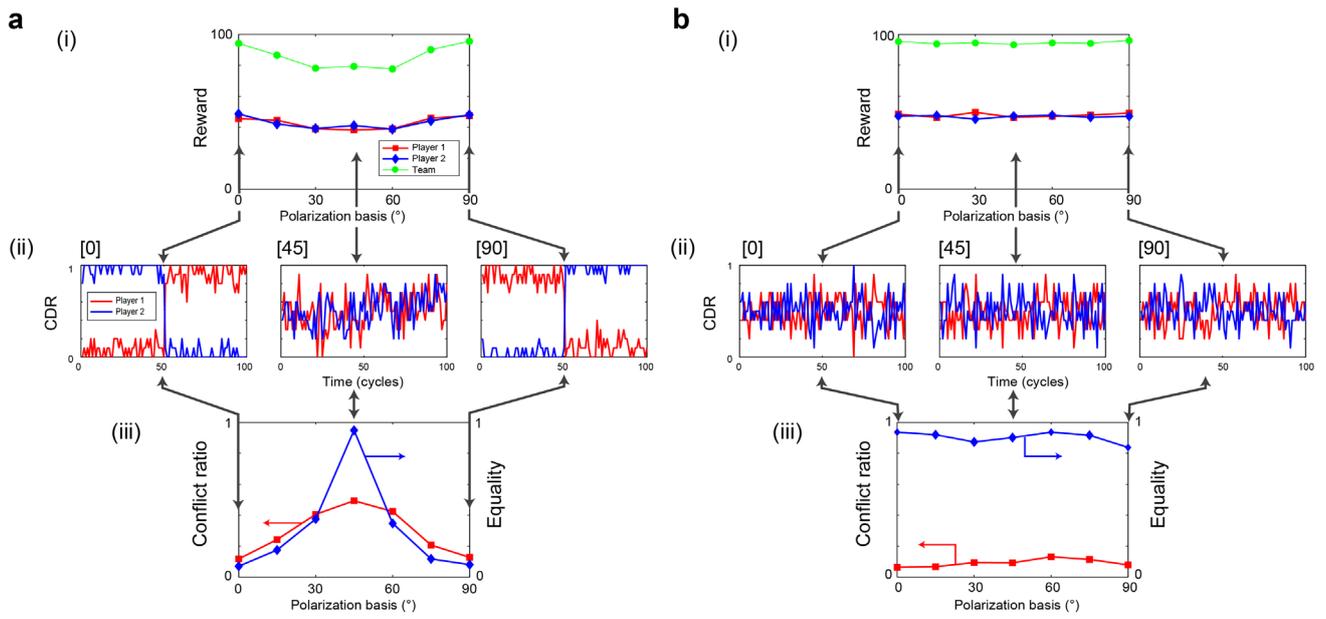

**Fig. 3 | Collective decision making using polarization-correlated and polarization-entangled photon pairs. a** Detailed analysis of the case of orthogonally polarized photon pairs. **b** Detailed analysis of the case of entangled photon pairs, where the low conflict ratio and high equality are preserved regardless of the polarization basis.

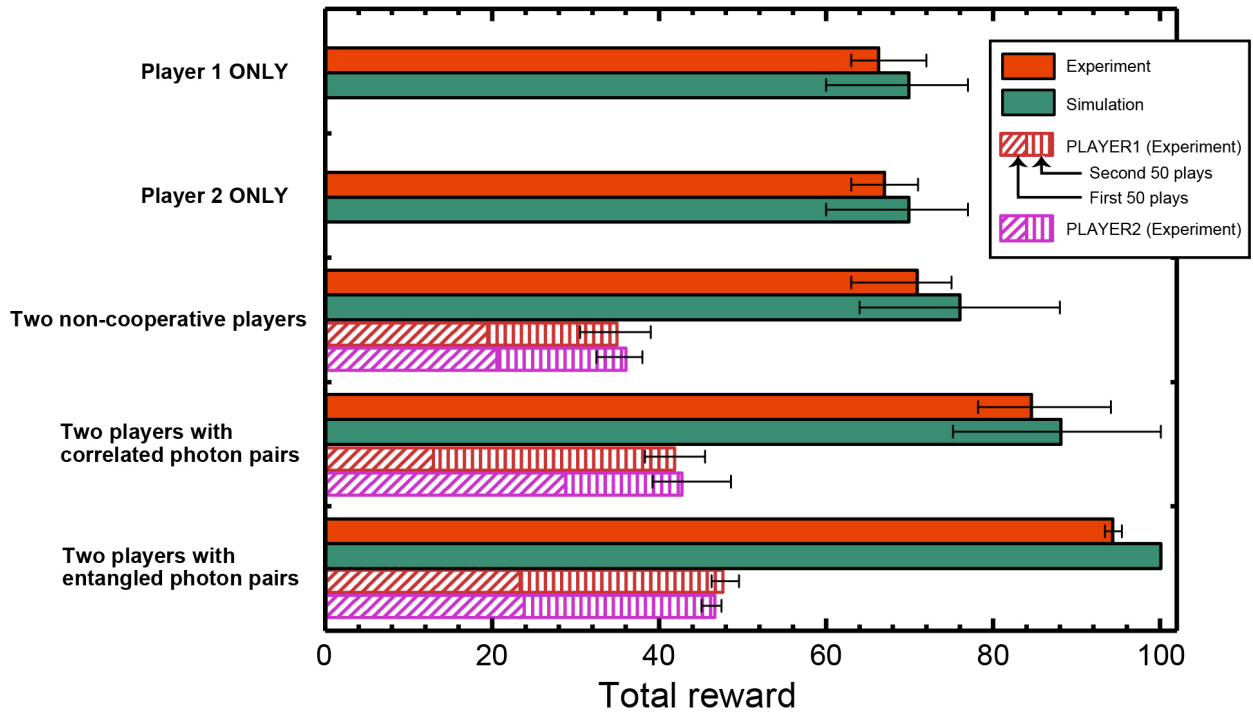

**Fig. 4 | Comparison of total rewards.** Comparison of the accumulated total reward after 100 plays, averaged over 10 repetitions, between the cases of a single player, two non-cooperative players, and two players with polarization-correlated and entangled photon pairs.

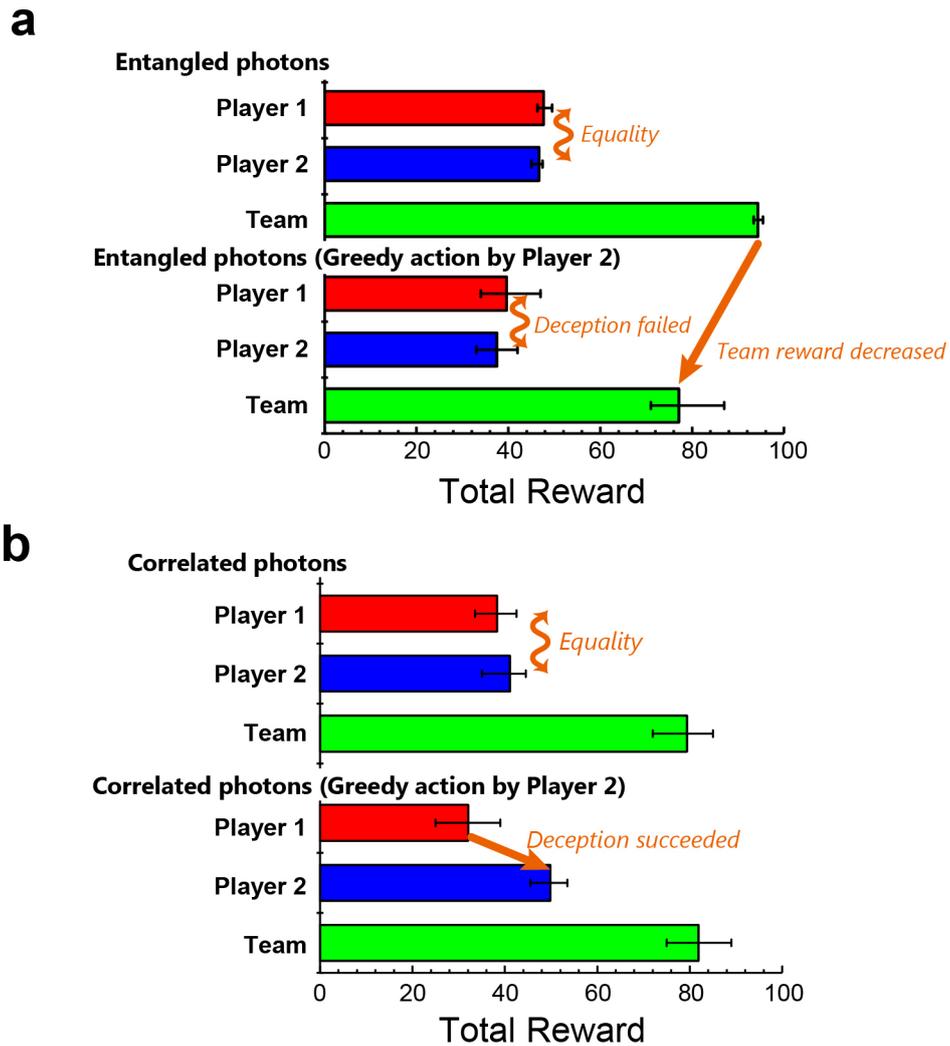

**Fig. 5 | Prevention of deception or greedy action.** Comparison of individual's and team's accumulated rewards at cycle 100. **a** With polarization-entangled photons, the accumulated reward of Player 2 is almost equal to that of Player 1, meaning that deception failed. Furthermore, the total team reward decreases. **b** With polarization-correlated photons, the reward accumulated by Player 2 is greater than that of Player 1, namely, deception is accomplished.

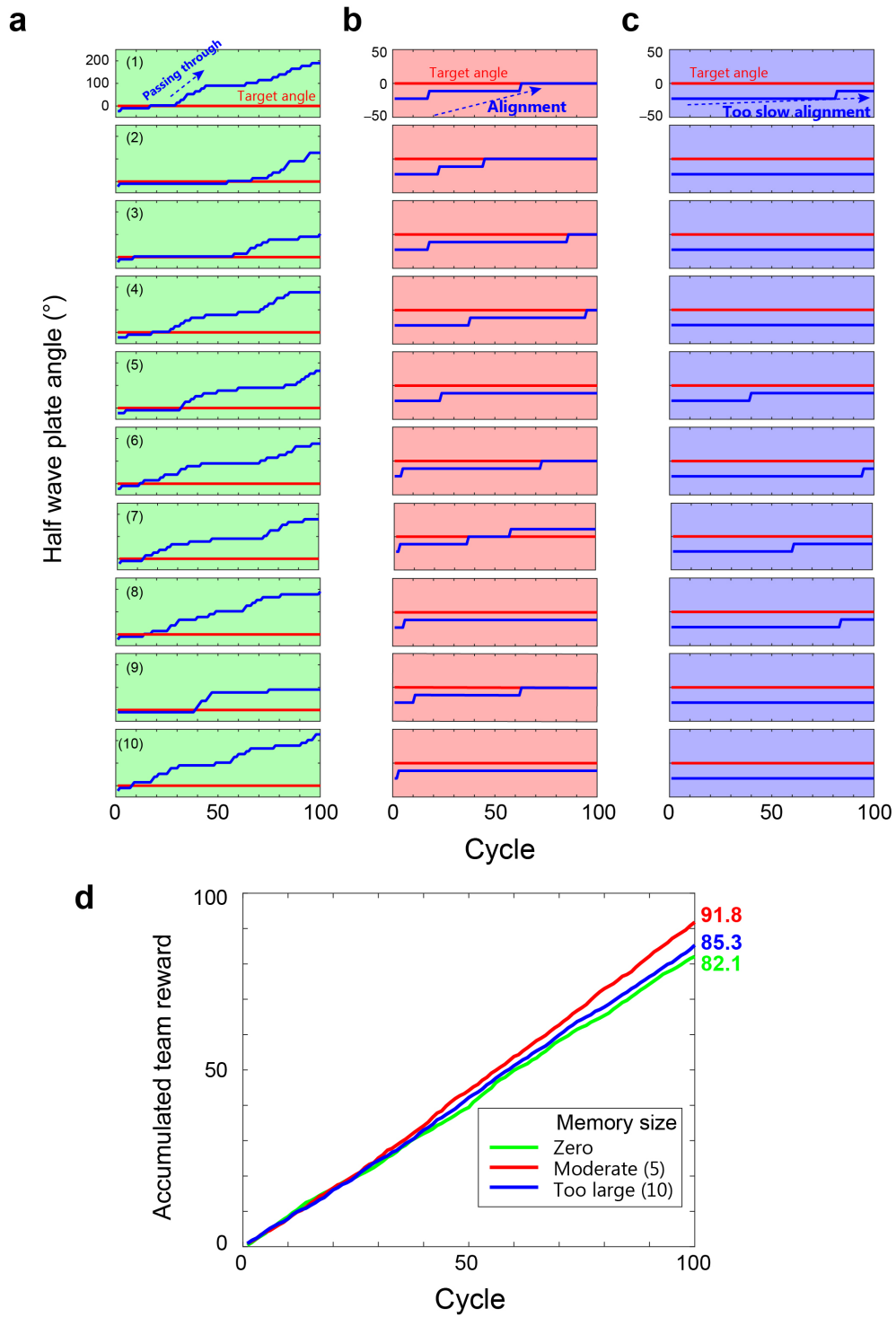

**Fig. 6 | Autonomous polarization basis alignment under Assumptions I and II.** Without any prior information, autonomous alignment should be possible by gradually rotating one of the half-wave plates. **a** However, due to the error signals that sometimes occur even when the polarization bases are aligned, the mechanism does not work well since the system *passes through* the optimal situation. **b** By referring to the recent history of the events involving decision conflict, robustness against errors is accomplished. **c** With too much reference to past events, the reaction becomes very slow.